\documentstyle[12pt,epsf]{article}
\begin{document}
\setlength{\headheight}{0in}
\setlength{\headsep}{0in}
\setlength{\topskip}{1ex}
\setlength{\textheight}{8.5in}
\setlength{\topmargin}{0.5cm}
\setlength{\baselineskip}{0.24in}
\catcode`@=11
\long\def\@caption#1[#2]#3{\par\addcontentsline{\csname
  ext@#1\endcsname}{#1}{\protect\numberline{\csname
  the#1\endcsname}{\ignorespaces #2}}\begingroup
    \small
    \@parboxrestore
    \@makecaption{\csname fnum@#1\endcsname}{\ignorespaces #3}\par
  \endgroup}
\catcode`@=12
\newcommand{\newc}{\newcommand}
\newc{\gsim}{\lower.7ex\hbox{$\;\stackrel{\textstyle>}{\sim}\;$}}
\newc{\lsim}{\lower.7ex\hbox{$\;\stackrel{\textstyle<}{\sim}\;$}}
\newc{\mtpole}{M_t}
\newc{\mbpole}{M_b}
\newc{\mtop}{m_t}
\newc{\mbot}{m_b}
\newc{\mz}{m_Z}
\newc{\mw}{m_W}
\newc{\alphasmz}{\alpha_s(m_Z^2)}
\newc{\MS}{\hbox{\rm\overline{MS}}}  \newc{\msbar}{\MS}
\newc{\DR}{\hbox{\rm\overline{DR}}}
\newc{\tbeta}{\tan\beta}
\newc{\stopL}{{\widetilde t_L}}
\newc{\stopR}{{\widetilde t_R}}
\newc{\stopone}{\widetilde t_1}
\newc{\stoptwo}{\widetilde t_2}

\newc{\BR}{\hbox{\rm BR}}
\newc{\xpb}{\hbox{\rm\, pb}}
\newc{\zbb}{Z\to b\bar}
\newc{\Gb}{\Gamma (Z\to b\bar b)}
\newc{\Gh}{\Gamma (Z\to \hbox{\rm hadrons})}
\newc{\rbsm}{R_b^\hbox{\rm sm}}
\newc{\rbsusy}{R_b^\hbox{\rm susy}}
\newc{\drb}{\delta R_b}
%
\newc{\xc}{\chi^{\pm}}
\newc{\xn}{\chi^{0}}
\newc{\swsq}{\sin^2\theta_W}
\newc{\tw}{\tan\theta_W}
\newc{\cw}{\cos\theta_W}
\newc{\sw}{\sin\theta_W}
\newc{\mhp}{m_{H^\pm}}
\newc{\mhalf}{m_{1/2}}
\newc{\ie}{{\it i.e.}}
\newc{\etal}{{\it et al.}}
\newc{\eg}{{\it e.g.}}
\newc{\kev}{\hbox{\rm\,keV}}
\newc{\mev}{\hbox{\rm\,MeV}}
\newc{\gev}{\hbox{\rm\,GeV}}
\newc{\tev}{\hbox{\rm\,TeV}}
\def\order#1{{\cal O}(#1)}
\def\mass#1{m_{#1}}
\def\Mass#1{M_{#1}}
\def\NPB#1#2#3{Nucl. Phys. {\bf B#1} (19#2) #3}
\def\PLB#1#2#3{Phys. Lett. {\bf B#1} (19#2) #3}
\def\PLBold#1#2#3{Phys. Lett. {\bf#1B} (19#2) #3}
\def\PRD#1#2#3{Phys. Rev. {\bf D#1} (19#2) #3}
\def\PRL#1#2#3{Phys. Rev. Lett. {\bf#1} (19#2) #3}
\def\PRT#1#2#3{Phys. Rep. {\bf#1} (19#2) #3}
\def\ARAA#1#2#3{Ann. Rev. Astron. Astrophys. {\bf#1} (19#2) #3}
\def\ARNP#1#2#3{Ann. Rev. Nucl. Part. Sci. {\bf#1} (19#2) #3}
\def\MODA#1#2#3{Mod. Phys. Lett. {\bf A#1} (19#2) #3}
\def\ZPC#1#2#3{Zeit. f\"ur Physik {\bf C#1} (19#2) #3}
\def\APJ#1#2#3{Ap. J. {\bf#1} (19#2) #3}
\def\MPL#1#2#3{Mod. Phys. Lett. {\bf A#1} (19#2) #3}
\def\beq{\begin{eqnarray*}}
\def\eeq{\end{eqnarray*}}
\def\bea{\begin{eqnarray*}}
\def\eea{\end{eqnarray*}}

%
\begin{titlepage}
\begin{flushright}
{\setlength{\baselineskip}{0.18in}
{\normalsize
hep-ph/9510372 \\
SLAC-PUB-7038\\
UM-TH-95-24\\
October 1995\\
}}
\end{flushright}
\vskip 2cm
\begin{center}
{\Large\bf Implications of the reported deviations from \\
         the standard model for $\Gamma (Z\to b\bar b)$ and $\alphasmz$}
\vskip 1cm

{\large
James D.~Wells${}^{a,}$\footnote{Work support by the Department of Energy,
contract DE-AC03-76SF00515.}  and G.L.~Kane${}^b$ \\}

\vskip 0.5cm
{\setlength{\baselineskip}{0.18in}
{\normalsize\it ${}^a$SLAC, MS 81  \\
     P.O. Box 4349 \\
     Stanford, CA 94309 \\}

\vskip 4pt
{\normalsize\it ${}^b$Randall Physics Laboratory \\
           University of Michigan \\
	   Ann Arbor, MI 48109--1120 \\}}

\end{center}
\vskip .5cm
\begin{abstract}

If the reported excess (over the standard model prediction) for
$Z\to b\bar b$ from LEP persists, and is explained by supersymmetric
particles in loops, then we show
that (1) a superpartner (chargino and/or stop) will be detected at
LEP2, and probably at LEP1.5 in 1995, (2) the basic parameter $\tbeta$ is
at its lower perturbative limit,
(3) $\BR (t\to \stopone+\tilde\chi^0)$ is at or above 0.4,
(4) the upper limit on $m_h$ is considerably reduced,
and (5) several important consequences arise for the form
of a unified supersymmetric theory.  Our analysis is done in
terms of a general weak scale Lagrangian and does not depend
on assumptions about SUSY breaking.

\end{abstract}
\end{titlepage}

\setcounter{footnote}{0}
\setcounter{page}{1}
\setcounter{section}{0}
\setcounter{subsection}{0}
\setcounter{subsubsection}{0}


\section*{Introduction}

For a year or so evidence has been getting stronger for two
deviations from the standard model.
One is an excess of about $2.1\pm 0.7\%$ in the $Z$ decays
to $b\bar b$ (using the value reported~\cite{denton:beijing,lepewwg95:02}
when the charm quark width
is fixed at
its standard model
value, for reasons explained below) denoted by $R_b$,
and the second
that the $\alphasmz$
measured from the $Z$ line width at LEP differs from that
determined other ways~\cite{shifman95:605,erler95:441,akhoury:private}.
For the first, the effect is
even larger~\cite{denton:beijing,lepewwg95:02}
if one uses unconstrained data ($2.73\pm 0.79\%$).
Also, these numbers
are for $\mtop =170\gev$; the deviation between theory and experiment
increases as $\mtop^2$
if one uses larger $\mtop$.
Experimentally these are logically independent deviations---for
example, if the excess $b \bar b$ were due to including charm decays in the
$b$ sample, there would be no effect on $\alphasmz$
since the total hadronic width
would be unchanged.  Theoretically they are also logically independent.
For example, if the predicted $\alpha_s$ in a model were lowered by
high scale threshold effects or intermediate scale matter multiplets,
there would be no necessary increase in $R_b$.
If these are true deviations they are the long-awaited
clues to physics beyond the standard model!

{}From a supersymmetric view these two deviations are natural
and expected.
The standard model
value for $R_b$ is the tree value minus about a $2\%$
effect from the $t-W^+$ loop (proportional to $\mtop^2$).  The corresponding
SUSY stop-chargino
loop naturally has the opposite sign of the $t-W^+$ loop, and
approximately cancels it if the stop and chargino are light enough.
Further, if the excess $Z$ decays are due to a new mechanism such as a
stop-chargino loop,
then this contribution must be included when the increase in the $Z$ width
is used to determine $\alpha_s$;
when that is done the $\alpha_s$ deviation also goes
away, and $\alpha_s$ from the $Z$ line width decreases to about .112,
consistent
with its determination other
ways~\cite{erler95:441,shifman95:605,kane95:350,chankowski95:304,garcia95:349}.
Thus the existence of the $\alpha_s$
deviation considerably strengthens one's confidence that both deviations are
real, and also that the SUSY explanation is perhaps correct.
It has been confirmed with global fits to all the precision
data~\cite{kane95:350,chankowski95:304,garcia95:349} that including
the SUSY contributions does not lead to disagreement with any
observable, and indeed that SUSY gives a better global fit to the
data than the standard model.

In this paper we argue that if the $R_b$ deviation is indeed real,
then several consequences follow.  (1) Most important, stop and chargino must
be light enough to be detected when the energy of LEP is increased to over
140 GeV, as expected during 1995, if sufficient luminosity is obtained (over
about $5\xpb^{-1}$).  To put it differently,
if a stop or chargino is not
found, then either the $R_b$ excess will go away,
or if it persists the SUSY
explanation is not relevant and there are different effects that
change $R_b$.  (2) By combining the $R_b$ effects with
other data we can show that $R_b$ can only be explained
in SUSY if $\tbeta$ is of
order 1.  Earlier arguments~\cite{boulware91:2054,wells94:219,garcia95:349,
chankowski95:304} that perhaps large $\tbeta$ and an $A-h$ loop
with small $m_A$ could also explain $R_b$ can be excluded.  (3) Since the stop
is lighter than the top there will be a decay of top to stop plus the
lightest superpartner (LSP). We show the branching ratio
for this decay must be large,
about 0.4. (4) The upper limit on  $m_h$ decreases considerably, making
its detection at LEP and/or FNAL more probable.

All of this analysis is essentially model and parameter
independent.  While
many of the relevant quantities depend on masses and couplings, we vary
them over all values allowed by constraints, and make no assumptions about
them.  The only assumption is that there are no other contributions except
those coming from standard model particles and their superpartners.  There is
no
dependence on the form of the theory at a high scale, on supersymmetry
breaking, etc., and no assumption of a MSSM (minimal supersymmetric
standard model)~\cite{mssm:note}.  We do feel that it is appropriate to
restrict
$\tbeta\gsim 1$ from perturbativity requirements on the top quark Yukawa
and other experimental considerations.   Also, we have not included
gluino/sbottom diagrams~\cite{djouadi93:3081} in our calculations
since we expect these to be
well below the neutralino/sbottom and chargino/stop
contributions.

The results do have implications for the form that models and the high
scale theory can take, and a fifth consequence (in addition to the 4 above)
is that there are at least three independent ways in which the high scale
theory must differ from the MSSM.  The MSSM is excluded if the $R_b$ excess
is true.

Greater than $2\sigma$ deviations from the standard model
have been reported for
$R_c$ and for $A_b(\equiv (g_L^2-g_R^2)/(g_L^2+g_R^2))$.
Supersymmetry has no natural way to explain these, and
the SUSY prediction is that they will go away.  Lest the reader think we
are arbitrarily
choosing our deviations, we note that for $R_c$ the reported deviation is
over a $10\%$ effect (compared to about $2\%$ for $R_b$ where
the $\mtop^2$ dependence
leads one to expect a large effect), and we are confident that no
mechanism could give such a huge effect without being detected other ways.
$R_c$ has no significant $\mtop^2$ contribution.  Thus any effect for $R_c$
should be well within
its current reported $~4\%$ errors, and that is why we quote
the data with $R_c$ constrained to its
standard model value above (which is conservative in any case).
For $A_b$ the errors
are a few $\%$ while the size of the expected effects in SUSY
is less than $1\%$.
Also, contributions to $R_b$ are sensitive to the left-handed couplings of $b$
to $Z$, while the asymmetry is most sensitive to the right-handed coupling,
which will be smaller.  In both cases the present errors are well above any
possible loop effect.

\section*{Stop and chargino at LEP}

Figure 1 shows the region of stop and chargino masses where a
$\drb$ of .003 or more can be obtained.  We think .003 is a good value to
use as a criterion to explain $\drb$.
The standard model gives $R_b=.216$ ($\mtop =170$),
and the reported $R_b$ value for charm constrained to its standard
model value is
$.2205\pm .0016$~\cite{denton:beijing,lepewwg95:02}, so
the $1\sigma$ lower limit is $0.003$ above the standard model.
Also, the effect on
the LEP $\alpha_s$ is
about $-4\drb$, so a change of $0.003$ in $R_b$
would yield a change of $-0.012$ in $\alpha_s$.  This is
exactly what is
needed to get the $\alphasmz$ extracted using the
$Z$ line shape down to about $0.112$ (from $0.124$)
where other ways of determining $\alpha_s$ lead us
to expect it.
The line in figure 1
is plotted for $\mtop =170$ GeV and for $\tbeta = 1.1$.  A rule is given in the
caption for scaling to other $\mtop$ and $\tbeta$.
This region is obtained by varying all parameters over values that do
not lead to a contradiction with theory or data, for each combination
of stop and chargino masses.  A point inside the region $\drb >0.003$
gives $\drb >0.003$ for some values of other parameters, though
not necessarily for all.  A point outside does not
give $\drb \geq 0.003$ for any parameter values.

The chargino cross-section is large enough at LEP so chargino pairs
could be copiously produced
nearly up to the kinematic limit~\cite{martin:private}
with several pb${}^{-1}$ of integrated luminosity
regardless of the values of other parameters.  Note that if the chargino is
not detected below about $70\gev$, then the stop mass should be lighter than
60 GeV.  Stop cross-sections are smaller, but with over $5{\xpb}^{-1}$
perhaps a few stop events could be detected~\cite{martin:private}.
Thus if the $R_b$ excess is real and
LEP takes data at or above 140 GeV and over about $5{\xpb}^{-1}$,
charginos and/or
stops can be detected if SUSY is relevant to understanding $R_b$!

We do not have space for a detailed analysis of chargino or stop
signatures at LEP, but a few points should be made.  We do not have any
firm arguments about whether stop or chargino is lighter.  If the chargino is
heavy enough to decay to $\stopone + b$, then the chargino pair final state
gives $b\bar b c\bar c\tilde\chi^0 \tilde\chi^0$, and stop pairs give
$c\bar c\tilde\chi^0\tilde\chi^0$; there are no leptons at all, though
the usual SUSY acoplanarity and acolinearity are present
($\tilde\chi^0$ is the LSP, and $\stopone$ the light stop mass
eigenstate).  As charginos get
lighter so that decay to $\stopone + b$ is excluded,
they should have a large branching ratio to a
charged lepton, albeit a rather soft one.
If $\stopone > \widetilde\chi^+ + b$, then the
chargino pairs give two charged leptons, two $\nu$'s, and two $\tilde\chi^0$'s,
or four jets plus two $\tilde\chi^0$'s,
while the stop pairs give $b\bar bl^+l^-\nu\nu\tilde\chi^0\tilde\chi^0$.

For much of the parameter space with sufficiently large $\drb$ we expect
that LEP1.5 will discover a chargino or stop.  However, it is possible
that the luminosity or center of mass energy will be too low to detect
them in some portions of parameter space.  Then LEP2 would be required.
Perhaps the optimal energy of LEP2 for these purposes
would be slightly less than
the $W^+W^-$ threshold, in order to reduce the
background to SUSY chargino events.

\section*{$\tan\beta$ is near 1}

Earlier
studies~\cite{denner91:695,boulware91:2054,wells94:219,garcia95:349,chankowski95:304}
have sometimes argued that a SUSY loop containing
the pseudoscalar ($A$) and scalar Higgs bosons ($h$ and $H$)
could, if $\tbeta$ were sufficiently large, give a large
contribution to $R_b$.  And when the contributions due to
chargino/stop and neutralino/sbottom loops are added, a significant
enhancement of $R_b$ is possible. However, there are a number of
constraints that must
be examined.  These include the decay
$Z\to b\bar bA(\to b\bar b)$~\cite{djouadi91:175} since the strength of
the $Ab\bar b$ vertex is proportional to $\tbeta$,
so if $\tbeta > 60$ the rate is
enhanced by over 3600;
$b\to c \tau \nu_\tau$~\cite{krawczyk88:182,grossman95:213};
and $Z\to \tau^+ \tau^-$, $t\to H^+ + b$,
$Z\to A+\gamma$, $b\to s+\gamma$.
It turns out that at the present time the first
two of these give the strongest constraints, and are sufficient to exclude
the large $\tbeta$ + small $m_A$ solution.  The others may in the future
strengthen this case, and data on all of these should be improved.

Figure 2 shows these constraints.  The approximately vertical lines
are shown labeled by the numbers of $Z\to b\bar b(A\to b\bar b$)
events that would have
been seen.  We are not yet aware of published data on this, but we think
that if 20-30 such events had been produced it is likely to have been
noticed, so we assume the allowed region in the $\tbeta$-$m_A$ plane is to
the right of those lines.  The calculation of~\cite{grossman95:213}
shows that the observed
rate for $b\to c\tau\nu_\tau$
requires that $\tbeta/M_{H+}<.52\gev^{-1}$.  This
gives a constraint on $m_A$---note that some model dependence
enters into this constraint
that could be changed if the Higgs sector were non-minimal in an unexpected
way.  The tree level constraint follows from the sum rule
$m_{H+}^2=m_A^2+m_W^2$.  This relation
must be radiatively corrected by loop effects, and the corrections are
large if $\tbeta$ is
large~\cite{gunion89:2333,berger90:225,diaz92:4246,brignole95:258}.
The results are shown as the approximately
horizontal line in the figure; the region above the line is excluded.  But
to explain $\drb$ the parameters must be in the
region above the .003 line,
which does not overlap with the allowed region.
It was shown previously~\cite{wells94:219,chankowski95:304,garcia95:349}
that to explain $\drb$ required either $\tbeta\simeq 1$ or very large.
Therefore SUSY can only
explain $\drb$ for $\tbeta$ about 1.

This has a number of important consequences.  It allows $b-\tau$
unification, but excludes $b-\tau-t$ unification of Yukawa
couplings.  The large top mass must be due to a
large top Yukawa coupling.  It is
consistent with an interesting explanation for the $\mu$
parameter~\cite{giudice88:480,brignole95:258}.  It
lowers the upper limit on the mass of the
lightest Higgs boson (see below).
It allows the LSP to have a large higgsino
component (as needed for $R_b$~\cite{wells94:219})
without disagreeing with the invisible width of the $Z$.

\section*{Light stop and top physics}

The LSP is lighter than the chargino.  Since the sum of our upper
limits on stop and chargino are less than $\mtop$, top will necessarily decay
to $\stopone + \tilde\chi^0$
with little kinematic suppression.  As discussed in
ref.~\cite{wells94:219} this lightest stop must be mostly
the superpartner of the right-handed top, and the light chargino and
neutralinos must be mostly Higgsino-like.  This, together with
the requirement that $\tbeta \sim 1$, largely determines the couplings
between the light stop and Higgsinos.  We find that the
branching ratio for $t\to \stopone + \tilde\chi^0_i$
($\tilde \chi^0_i$ is mostly LSP) is larger
than 0.4 for all allowed
choices of parameters, so it should be seen at FNAL once
it is looked for.

If stop is heavier than chargino it will decay to $\tilde\chi^+ + b$; if
not, to $c+ \tilde\chi^0$.  We do not have room
here to go through signatures and
a detailed analysis, but we note several points.  CDF has published a
branching ratio
of $.87^{+.18}_{-.32}$~\cite{cdf95:237}
for $t\to W^+b$.  But that analysis was for a
sample with a $W$ leptonic decay trigger, so it does not apply to
some events with
a decay to $c + \tilde\chi^0$, and without detailed analysis
it is not clear what
fraction of decays to stop could pass if
$\stopone\to \tilde\chi^+\to ``W"+\hbox{\rm LSP}$.
In any case, we think a branching ratio of 0.4 is not
excluded~\cite{mrenna95:424}.
We are aware
that a smaller branching ratio for $W^+b$
implies a larger production cross-section for top,
and therefore a smaller mass.
This situation is interesting, and we are not
quite sure about its implications.
We note that if stop decays to
$c + \tilde\chi^0$ then the $t\bar t$ final state
will often have
$t(\to W + b)$ + $\bar t (\to \stopone (\to c + \tilde\chi^0)+\tilde\chi^0)$
so it will have extra $Wjj$ events.  A mild indication
of such an effect has been
reported~\cite{cdf95:2626}.

D0 has reported~\cite{white:susy95} some limits on light stops.
It is difficult to show the impact of this on Figure 1 without model
dependence since one has to
relate charginos and neutralinos.  For stops above about 60 GeV it
constrains possible solutions, but has little impact below that.  This
could be interpreted as an argument for lighter stops, but is not
conclusive.

\section*{Prediction for $m_h$}

In SUSY the value of the lightest Higgs boson mass can be
calculated, but it depends on other parameters.  There is an
upper limit, independent of models, of about
$150\gev$~\cite{kane93:2686,espinosa93:51}.
The MSSM gives upper limits
of about $130\gev$.
In all cases $m_h$ has a tree level value plus a large
contribution mainly from
top dependent one--loop corrections.  The tree
level limit is $\mz|\cos 2 \beta|$, so for $\tbeta$ near 1 this is very small,
and the upper limit on $m_h$ is considerable reduced.  In the
minimal model it is then well below 100 GeV.
The loop contributions are also reduced when the
stop mass is small.  Therefore it is nearly certain that
LEP will find $h$ (if $\drb$ is real and explained by SUSY) if a total
energy over 190 GeV and $500{\xpb}^{-1}$
are obtained.  With $\tbeta$ near 1 the
light Higgs is rather standard-model-like,
but even if the $Zh$ cross-section were suppressed
the $Ah$ cross-section should be large enough.

\section*{Consequences for theory}

There are at least three major consequences of these arguments for
the form a supersymmetric unified theory can take.  Two of them
have been
remarked on before~\cite{kane95:350}
and we only briefly comment on them here.  Each of
the three excludes the MSSM.  It is exciting that data at the
electroweak scale may be constraining the theory at the unification
scale -- if it is, once we have information on superpartner properties
it may be possible to determine much of the effective Lagrangian
at the unification scale from experiments at the electroweak scale.

It is well known that in order to have light superpartners in the
MSSM it is necessary to have $\alphasmz > .126$~\cite{bagger95:443},
while we see here that $\alphasmz$ is
about .112.  Thus the theory must have some additional
structure in order to lower $\alphasmz$.

Second, the stop and chargino mass matrices and couplings must be
such that the chargino is largely higgsino and the stop mainly right-handed,
in order that the $\stopone-\tilde\chi^+-b$ vertex
(which is proportional to $\mtop$)
enter at full strength while the chargino and stop are light, and other
constraints are met.  That cannot happen in the MSSM when one looks
carefully at the conditions~\cite{wells94:219}.

Third, we have seen here that we require a value of $\tbeta$ about
1.1.  That is lower than the perturbative lower limit given
approximately by $\sin\beta> \mtop/200\gev$.
One could view that as evidence against a SUSY explanation of
$\drb$, but we think
that is premature, because the form of a more complete supersymmetric
theory can
affect the running of the top Yukawa, thus lowering the
allowed $\tbeta$.  We prefer to view it as a constraint
on the form of a satisfactory unified supersymmetric theory, a constraint
that the MSSM fails.

{\setlength{\baselineskip}{0.1666in}

\section*{Acknowledgements}

We have benefitted from conversations with and information from
A.~Brignole, C.~Baltay, D.~Gerdes, H.~Haber, S.-B. Kim, G.~Kribs,
P.~Langacker, S.~Martin, and A.~White.
J.D.W. thanks the LBNL theory group for their kind support
during a recent visit.  G.L.K. thanks the Insitute
for Theoretical Physics, Santa Barbara, where part of this
work was done, for its hospitality and
support. G.L.~Kane is supported in part
by the U.S. Department of Energy.

\section*{Figure Captions}

\begin{itemize}

\item Figure 1: Contour of $\drb=0.003$ in the
$m_{\chi^\pm_1}-m_{\tilde t_1}$
plane with $\mtop=170\gev$ and $\tbeta=1.1$.  Above the contour no solution
exists which yields $\drb>0.003$.  Below the contour solutions do exist
with $\drb >0.003$ for appropriate choices of parameters.
The numerical value of this contour is enhanced
(or diminished) by about $(0.4/\sin\beta)^2(\mtop/\mz)^2$
for different choices of $\mtop$ and $\tbeta$.

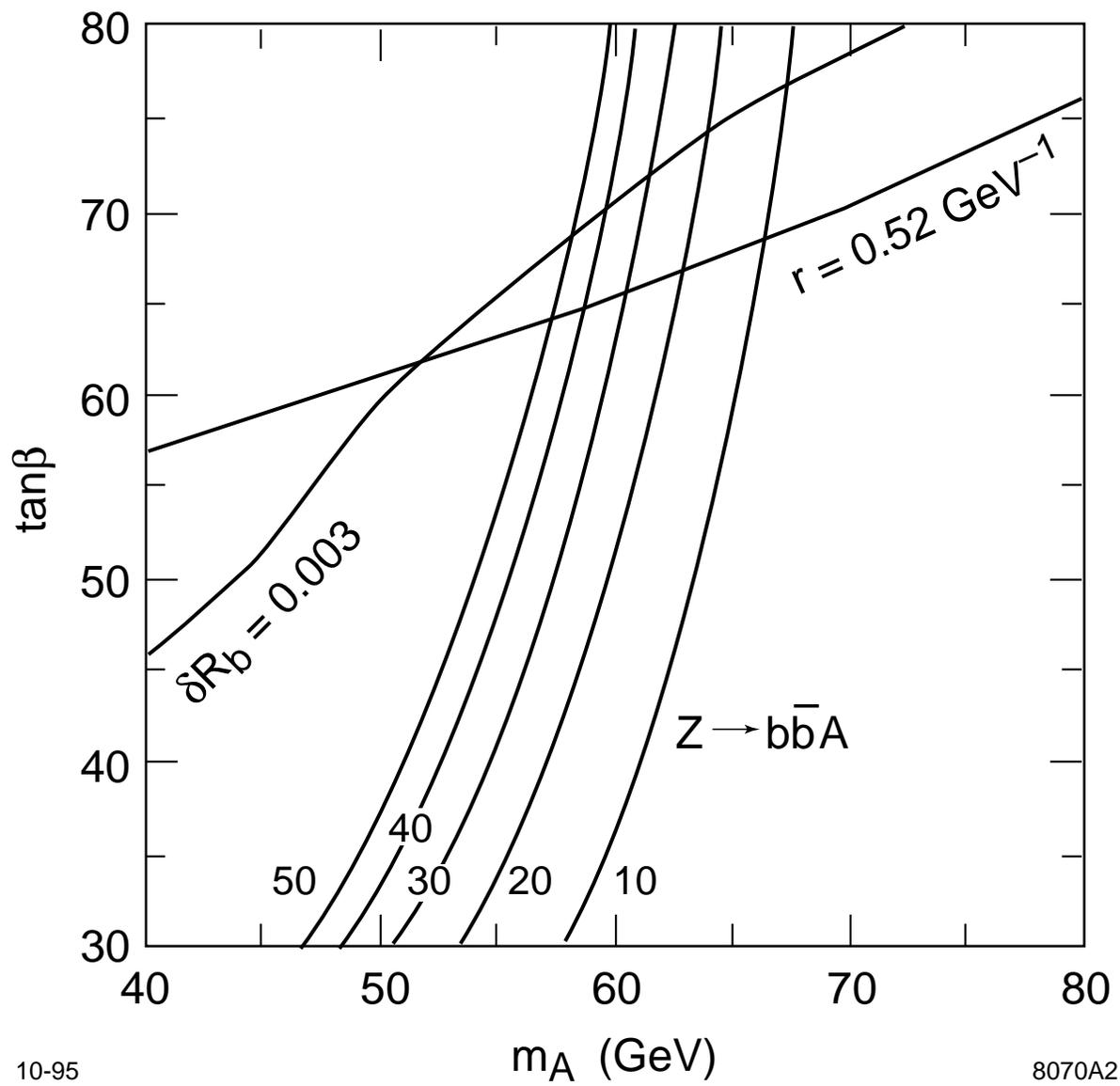
\item Figure 2: The high $\tbeta$ exclusion plot.  The $\drb =0.003$
contour is plotted such that no supersymmetric
solution below the contour can provide $\drb\geq 0.003$.
The region above the $r=0.52\gev^{-1}$ contour is
excluded by $b\to c\tau\nu_\tau$
decay data.  The region to the left of the vertical lines,
which indicate contours
of $Z\to b\bar b A$ events, is also excluded.  Therefore, if we require
$\drb >0.003$, which we argue for in the text, then no region
of parameter space is simultaneously
consistent with the $b\to c\tau\nu_\tau$ and
$Z\to b\bar bA$ decay constraints.

\end{itemize}

}

\begin{thebibliography}{99}
\bibitem{mssm:note}
	By MSSM (Minimal Supersymmetric Standard Model) we are
referring to a supersymmetric model with minimal particle content,
simple gauge coupling unification at the high scale, common
scalar masses and gaugino masses at the high scale.  It is
equivalent to the model studied in G.~Kane, C.~Kolda, L.~Roszkowski,
J.~Wells, \PRD{49}{94}{6173}.
\end{thebibliography}
\end{document}